\documentclass[12pt]{article}

\usepackage{scicite}

\usepackage{times}
\usepackage{color}
\usepackage{graphicx,amsmath}
\usepackage{epstopdf}
\usepackage {indentfirst}
\usepackage{float}

\newenvironment{sciabstract}{%
\begin{quote} \bf}
{\end{quote}}

\title{\Large Near-symmetric multiport beam splitting for high-NOON state preparation on nonlocal metasurface}

\author
{Yu Tian,$^{1,2}$ Qi Liu,$^{1,2}$ Zhaohua Tian,$^{1}$ Qihuang Gong,$^{1,2,3,4,5}$ and Ying Gu,$^{1,2,3,4,5,\ast}$\\
\\
\small{$^{1}$State Key Laboratory for Mesoscopic Physics, Department of Physics, Peking University, Beijing 100871, China}\\
\small{$^{2}$Frontiers Science Center for Nano-optoelectronics $\&$ Collaborative Innovation Center of QuantumMatter $\&$} \\ 
	\small{Beijing Academy of Quantum Information Sciences, Peking University, Beijing 100871, China}\\
\small{$^{3}$Collaborative Innovation Center of ExtremeOptics,Shanxi University,Taiyuan,Shanci 030006,China}\\
\small{$^{4}$Peking University Yangtze Delta Institute of Optoelectronics, Nantong 226010, China}\\
\small{$^{5}$Hefei National Laboratory, Hefei 230088,China}\\
\small{$^\ast$ E-mail: ygu@pku.edu.cn}
}

\begin{document} 


\baselineskip24pt


\maketitle


\begin{sciabstract}
	Polarization beam splitting (BS) has been implemented on gradient metasurface with local response for entanglement manipulation and state reconstruction. To realize more degrees of light modulation, nonlocal modes, manifested as wavelength and momentum selectivity, should be applied into metasurface BS. Here, we demonstrate that single nonlocal phase gradient metasurface (NPGM) can function as a series of independent near-symmetric multiport BS, constructed by its momentum-polarization mode subspaces.Then, using any of above BS with simultaneous multiphoton interference, high-photon NOON states are prepared  with high success probability and fidelity. For example, four-mode four-photon  NOON state is obtained with 34.8\% success probability and fidelity of 99.9\%, greatly higher than those previously reported. With unique capability of multiphoton interference, this multiport BS on single NPGM can be directly used in the on-chip quantum photonics. Also, the efficient generation of high-photon NOON states with above BS has potential applications in quantum precision measurement.
  
\end{sciabstract}


\section*{Introduction}

Metasurfaces, as a kind of structed two-dimensional materials, possess the capability to fully control light within a subwavelength layer, which have been applied in the wavefront control and quantum state engineering ~\cite{1,2,3}.Among them, gradient metasurfaces can perform multiple functions of beam splitting (BS) with various split ratios, which splits the incident light into several beams according to the polarization or wavelength ~\cite{4,5,6,7,8,9,10,11,12,13,14}. Especially, polarization-dependent metasurface BS has exhibited various functions in the quantum state manipulation, such as quantum state reconstruction~\cite{5}, entanglement and disentanglement of two-photon spin states~\cite{6} and multichannel entanglement distribution/transformation, ~\cite{8}, etc. Compared with previous BS, metausrface BS has the advantages of small size, easy integration, flexible performance, which is significant for integrated photonic devices.However, most current researches are focused on BS of gradient metasurfaces with local response, lacking the utilization of momentum of light, which restricts the multifunctionality of metasurface BS.

Nonlocal meatsurfaces, working under nonlocal modes with spectral and angular selectivity, have been widely researched in the image processing and nonlinearity enhancement~\cite{15,16,17,18,19,20,21}. An important class of nonlocal metasurfaces is working under bound states in the continuum (BICs), which mostly focuses on enhancing light confinement~\cite{22,23,24,25,26,27} and BICs-assisted sensing~\cite{28,29,30}. Recently, a new kind of BIC-based metasurface with phase gradient, so called nonlocal phase gradient metasurfaces (NPGM), is proposed to further enhance the spatial modulation ability of nonlocal metasurfaces for light~\cite{31,32,33,34,35,36,37,38}. Nevertheless, the BS of nonlocal metasurface has received little attention, which ought to enable metasurface BS with the control of spectrum and momentum.
Therefore, one of the purposes of our work is to construct more functional BS on the NPGM through combining the advantages of gradient metasurface and nonlocal metasurface, which has potential to control multiple freedoms of light simultaneously to realize flexible and integrated manipulation of quantum states, such as, NOON states.

NOON states, having the form like $\frac{1}{\sqrt{2}}(|N_A0_B\rangle+|0_AN_B\rangle)$, as a kind of maximally N-photon entangled states, are utilized in the quantum precision measurement~\cite{39,40,41}.The major challenge for the NOON states is the generation of NOON states with high photon numbers, especially for the NOON states with N $\geq$ 3. Till now, high-photon NOON states have been prepared through multiphoton interference of classical and quantum states generated from spontaneous parametric down-conversion~\cite{42}, manipulation of polarization-encoded photons~\cite{43}, and integrated photonic circuits~\cite{44}, etc. Nevertheless, the current schemes for high-photon NOON states preparation requires a complex optical system with a relative low success probability, consisting of a set of two-port beam splitters , which restricts the practical availability of these NOON states. Instead, using NPGM, which supports multiport BS, will reduce the number of beam splitters in the system of high-photon NOON states preparation and improve multiphoton interference efficiency. Hereby, another purpose of our work is to employ the NPGM in NOON states preparation, reveal its progress in  success probability and fidelity of high-photon NOON states integrated preparation.

Here, we theoretically investigate the multiport BS functions of NPGM for multiphoton interference to prepare high-photon NOON states. 
The results indicate that there are a series of independent near-symmetric four-port  BS on single NPGM, which has different working wavelengths, enabling four-photon interference simultaneously. 
These BS, working under circular basis, are achieved through four- and eight-dimensional momentum-polarization mode (MPM) subspaces of NPGM. 
Taking advantages of these BS, four-mode high-photon NOON states can be obtained with high success probability and fidelity from initially non-entangled photons through multiphoton interference.
For example, we obtain two-photon NOON state with 47\% success rate, three-photon NOON state with 36\% success rate and four-photon NOON state with 34\% success rate, where their fidelity all reaches to 99\%. 
Our results indicate the advantages of multiport BS of nonlocal metasurfaces in the multiphoton interference and the manipulation of multiphoton quantum states, especially the preparation of NOON states. Moreover, the preparation of high-photon NOON states with high success rate and integration also promotes the development of quantum precision measurement and quantum sensing.

\section*{RESULTS}

\subsection*{Multiport BS in the MPM subspaces of NPGM}

\begin{figure}[ht]
	\centering
	\includegraphics[width=0.9\textwidth]{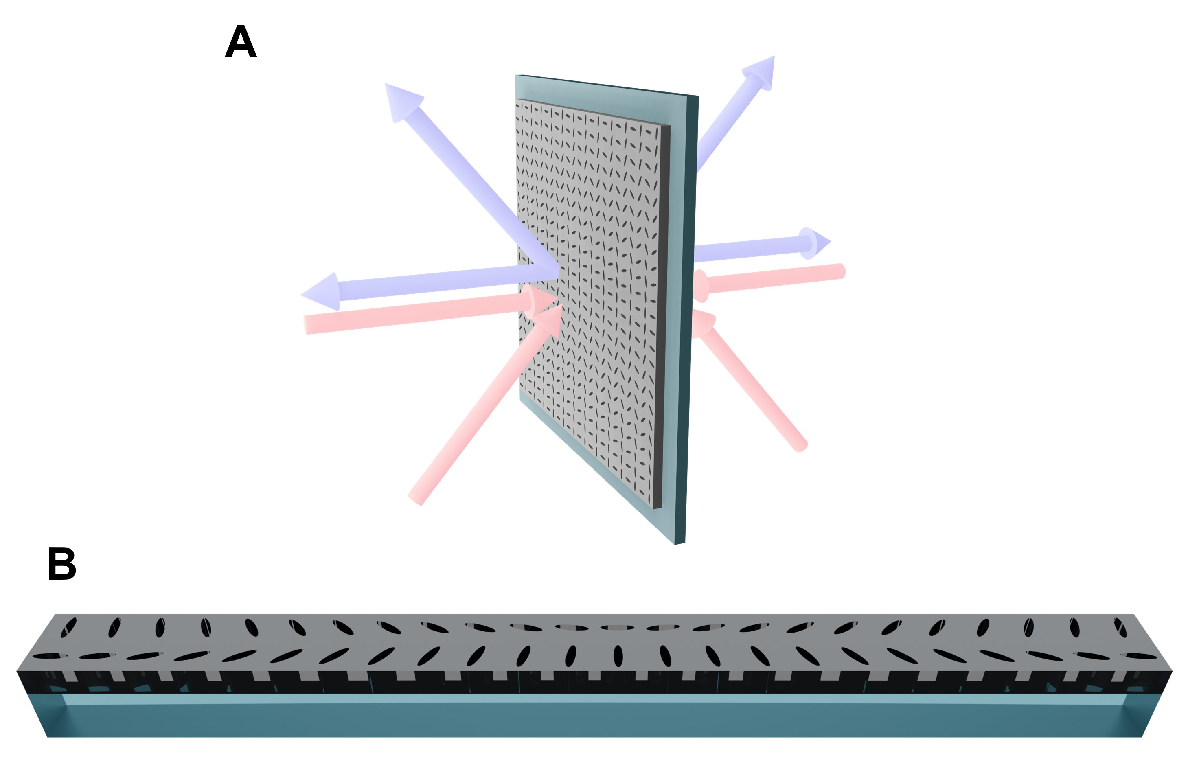}
	\caption{Schematic of multiport BS on NPGM. (A) Four beams of light with specific wavelength, momentum and polarization are near symmetrically converted into another four beams with specific wavelength, momentum and polarization after passing through NPGM. (B) Schematic of a period of meta atoms composed of a slab of silicon (the gray part) with ellipses holes on the glass substrate (the cyan part).}	
\end{figure}

Consider a metasurface consisting of a slab of silicon etched with elliptical holes periodically on the glass substrate [Fig. 1A]. 
Its arrangement of meta atoms in one period is shown in Fig. 1B, which is encoded with a linear geometric phase gradient (See supplementary Materials S1). 
Recent researches have shown that such a metasurface, known as NPGM, can support quasi BICs in the near infrared wavelengths, which can deflect the normal incident light into $\pm$2nd diffraction orders of the metasurface at the resonant frequency while the non-resonant light will not deflect~\cite{31,32,33,34,35,36,37,38}.
For obliquely incident light, there are two similar quasi BICs with right-circular polarization (RCP) or left-circular polarization (LCP) eigenstate, which means it can only be excited by RCP or LCP light under a specific wavelength, while the light with opposite chirality  cannot excite quasi BICs under this wavelength.
Therefore, the spectral positions of quasi BICs are determined by the momentum and polarization of incident light simultaneously.
By varying the momentum and polarization of incident light, we obtain four dispersion curves of NPGM for the same mode [Fig. 2A], corresponding to four different excite conditions of quasi BICs, respectively.
For different excite conditions, the distributions of electric field are similar, like Fig. 1B shows.

From the dispersion relationship, we can get a series of momentum-polarization mode (MPM) subspaces on the NPGM. 
Fig. 2A reveals that under a specific wavelength, there are many different excite conditions of quasi BICs for the NPGM, which are different in the polarization or momentum.
That is, there is a fixed group of MPM to excite quasi BICs with the specific wavelength,which have four ($\lambda_1$) or eight MPMs ($\lambda_2$ and $\lambda_3$).
These four or eight modes construct a series of MPM subspaces, where they can be converted to each other through NPGM.
On the other hand, our system is linear, which doesn't allow the wavelength conversion. 
Therefore, the transformation among MPMs at different wavelengths is impossible, revealing that those MPM subspaces are independent with each other, according to the working wavelengths, as well as the transformation rules of NPGM. 
These subspaces are four- or eight- dimensional, where the conversion of MPMs in the subspaces is possible while the conversion of MPMs among different subspaces is not allowed.
The features of these MPM subspaces indicate the potential of independent multiport BS with strict selectivity of wavelength, momentum and polarization on the NPGM.

For the subspace I shown in Fig. 2A, it corresponds to the only four-dimensional subspace which has maximum working wavelength of 1525.1 nm. The paths of light locate at -1st and +1st diffraction orders of NPGM. As shown in Fig. 2B,  there are four MPMs at four paths in this subspace, including $|L\rangle^d_{-1}$, $|R\rangle^d_{+1}$, $|L\rangle^a_{+1}$ and $|R\rangle^a_{-1}$, in which L/R refers to the polarization of light, d/a refers to the light propagate in glass or air side and -1/+1 refers to the path of light. The transformation of these four modes through NPGM can be equivalent to the four-port BS under specific working frequency, where the input and output modes are totally the same, both in the momentum and polarization, just like Fig. 2B shows.

\begin{figure}[H]
	\centering
	\includegraphics[width=\textwidth]{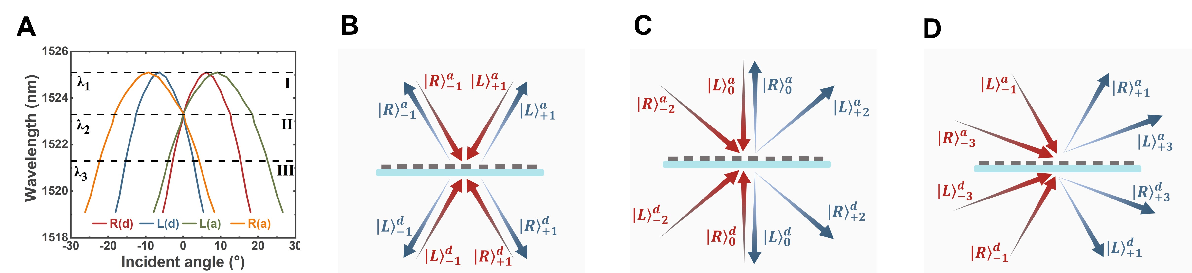}
	\caption{Momentum-polarization mode subspaces of NPGM and four-port beam splitting based on that. (B) The momentum-polarization mode transformation in the subspace I at $\lambda_1$. (C) The momentum-polarization mode transformation in the subspace II at $\lambda_2$. (D) The momentum-polarization mode transformation in the subspace III at $\lambda_3$.}	
\end{figure}

Considering the requirements of symmetry of NPGM, there are at most two independent BS processes in subspace I, whose amplitude and phase parameters are $S_0$/$\phi_0$ and $S_1$/$\phi_1$, respectively. Therefore, the BS processes for this subspace can be described as 
\begin{equation}
	\renewcommand{\arraystretch}{1.5}
{\left( {\begin{array}{*{20}{c}}
			{|L\rangle_{ - 1}^d}\\
			{|R\rangle_{ + 1}^d}\\
			{|L\rangle_{ + 1}^a}\\
			{|R\rangle_{ - 1}^a}
	\end{array}} \right)_{out}} = \left( {\begin{array}{*{20}{c}}
		{S_0^4{e^{i\phi _0^4}}}&{S_0^3{e^{i\phi _0^3}}}&{S_1^1{e^{i\phi _1^1}}}&{S_1^2{e^{i\phi _1^2}}}\\
		{S_0^3{e^{i\phi _0^3}}}&{S_0^4{e^{i\phi _0^4}}}&{S_1^2{e^{i\phi _1^2}}}&{S_1^1{e^{i\phi _1^1}}}\\
		{S_0^1{e^{i\phi _0^1}}}&{S_0^2{e^{i\phi _0^2}}}&{S_1^4{e^{i\phi _1^4}}}&{S_1^3{e^{i\phi _1^3}}}\\
		{S_0^2{e^{i\phi _0^2}}}&{S_0^1{e^{i\phi _0^1}}}&{S_1^3{e^{i\phi _1^3}}}&{S_1^4{e^{i\phi _1^4}}}
\end{array}} \right){\left( {\begin{array}{*{20}{c}}
			{|L\rangle_{ - 1}^d}\\
			{|R\rangle_{ + 1}^d}\\
			{|L\rangle_{ + 1}^a}\\
			{|R\rangle_{ - 1}^a}
	\end{array}} \right)_{in}},
\end{equation}
in which $S_{0/1}^{1}$/$\phi_{0/1}^{1}, $ $S_{0/1}^{2}$/$\phi_{0/1}^{2}$, $S_{0/1}^{3}$/$\phi_{0/1}^{3}$ and $S_{0/1}^{4}$/$\phi_{0/1}^{4}$ are real numbers, referring to the amplitude parameters and phase factors of the processes of normal transmission, abnormal transmission, normal reflection and abnormal reflection, respectively. Owing to the requirements of unitarity, the amplitude parameters must satisfy $\sum_{i=1}^4(S_{0/1}^i)^2=1$, which makes the number of free parameters is smaller than 16 we list. Besides, the limitations of nonlocal modes make all the amplitude parameters $S\approx0.5$. Thus, an independent near-symmetric four-port beam splitter is constructed on the NPGM with the selectivity of wavelength, polarization and momentum of input and output light. 

It is easy to get the specific description $M_1$ of the above beam splitting through commercial COMSOL Multiphysics after excluding the influence of propagation phases, which can be expressed as 
\begin{equation}
	\setlength{\arraycolsep}{2.5pt}
	\renewcommand{\arraystretch}{1.5}
{\left( {\begin{array}{*{20}{c}}
			{|L\rangle_{ - 1}^d}\\
			{|R\rangle_{ + 1}^d}\\
			{|L\rangle_{ + 1}^a}\\
			{|R\rangle_{ - 1}^a}
	\end{array}} \right)_{out}} = \left( {\begin{array}{*{20}{c}}
		{0.57{e^{ - i \times 74^\circ }}}&{0.45{e^{ - i \times 51^\circ }}}&{0.49{e^{ - i \times 92^\circ }}}&{0.48{e^{i \times 82^\circ }}}\\
		{0.45{e^{ - i \times 51^\circ }}}&{0.57{e^{ - i \times 74^\circ }}}&{0.48{e^{i \times 82^\circ }}}&{0.49{e^{ - i \times 92^\circ }}}\\
		{0.50{e^{ - i \times 92^\circ }}}&{0.48{e^{i \times 81^\circ }}}&{0.41{e^{ - i \times 126^\circ }}}&{0.59{e^{ - i \times 106^\circ }}}\\
		{0.48{e^{i \times 81^\circ }}}&{0.50{e^{ - i \times 92^\circ }}}&{0.59{e^{ - i \times 106^\circ }}}&{0.41{e^{ - i \times 126^\circ }}}
\end{array}} \right){\left( {\begin{array}{*{20}{c}}
			{|L\rangle_{ - 1}^d}\\
			{|R\rangle_{ + 1}^d}\\
			{|L\rangle_{ + 1}^a}\\
			{|R\rangle_{ - 1}^a}
	\end{array}} \right)_{in}}.
\end{equation}
The simulated results further demonstrate the four-port BS on the -1st and +1st diffraction orders of NPGM, whose amplitude parameters are close to 0.5. Despite that, we can still adjust concrete parameters of $M_1$ to a certain extent, especially the phase factors by changing the arrangement directions and geometric parameters of meta atoms.

For the subspace II shown in Fig. 2A, it corresponds to a special case for the eight-dimensional subsapces located at 1523.3 nm , which includes the condition of normal incident. In this subspace, the paths of light locate at -2nd, 0th and +2nd diffraction orders of NPGM. As shown in Fig. 2C, there are eight MPMs located at six different paths, including $|L\rangle_{-2}^d$, $|R\rangle_0^d$, $|L\rangle_0^a$, $|R\rangle_{-2}^a$, $|R\rangle_{ + 2}^d$, $|L \rangle_0^d$, $|R\rangle_0^a$, $|L\rangle_{ + 2}^a$. These eight modes can be divided into two parts: if we choose $|L\rangle_{ - 2}^d,|R\rangle_0^d,|L\rangle_0^a,|R\rangle_{-2}^a$ as input modes, then the output modes will be the combination of $|R\rangle_{ + 2}^d,|L \rangle_0^d,|R\rangle_0^a,|L\rangle_{ + 2}^a$ [Fig. 2C] and vice verse. The transformation of these eight modes can also be equivalent to the four-port BS, where the input and output modes are completely different, but part of them are overlap in the paths.

Compared with the previous BS, there are at most four independent BS processes in this subspace, whose parameters are labeled as $J_0$/$\varphi_0$, $J_1$/$\varphi_1$, $J_2$/$\varphi_2$, $J_3$/$\varphi_3$, respectively. Taking the BS process shown in Fig. 2C, we can describe the BS as  
\begin{equation}
	\renewcommand{\arraystretch}{1.5}
	{\left( {\begin{array}{*{20}{c}}
				{|R\rangle_{ + 2}^d}\\
				{|L\rangle_0^d}\\
				{|R\rangle_0^a}\\
				{|L\rangle_{ + 2}^a}
		\end{array}} \right)_{out}} = \left( {\begin{array}{*{20}{c}}
			{J_0^3{e^{i\varphi _0^3}}}&{J_1^4{e^{i\varphi _1^4}}}&{J_2^2{e^{i\varphi _2^2}}}&{J_3^1{e^{i\varphi _3^1}}}\\
			{J_0^4{e^{i\varphi _0^4}}}&{J_1^3{e^{i\varphi _1^3}}}&{J_2^1{e^{i\varphi _2^1}}}&{J_3^2{e^{i\varphi _3^2}}}\\
			{J_0^2{e^{i\varphi _0^2}}}&{J_1^1{e^{i\varphi _1^1}}}&{J_2^3{e^{i\varphi _2^3}}}&{J_3^4{e^{i\varphi _3^4}}}\\
			{J_0^1{e^{i\varphi _0^1}}}&{J_1^2{e^{i\varphi _1^2}}}&{J_2^4{e^{i\varphi _2^4}}}&{J_3^3{e^{i\varphi _3^3}}}
	\end{array}} \right){\left( {\begin{array}{*{20}{c}}
				{|L\rangle_{ - 2}^d}\\
				{|R\rangle_0^d}\\
				{|L\rangle_0^a}\\
				{|R\rangle_{ - 2}^a}
		\end{array}} \right)_{in}}.
\end{equation}
In this transformation matrix, $J_{0\sim3}^1$/$\varphi_{0\sim3}^1$, $J_{0\sim3}^2$/$\varphi_{0\sim3}^2$, $J_{0\sim3}^3$/$\varphi_{0\sim3}^3$, $J_{0\sim3}^4$/$\varphi_{0\sim3}^4$  are real numbers, referring to the amplitude parameters and phase factors of the processes of normal transmission, abnormal transmission, normal reflection and abnormal reflection, respectively. Similarly, due to the requirements of unitarity $\sum_{i=1}^{4}(J_{0/1/2/3}^i)^2$=1, the number of free parameters, including amplitude parameters and phase factors, is less than 32 we list. Besides, the amplitude parameters of this BS are also close to 0.5 because of the nonlcoal modes. Therefore, we can obtain another near-symmetric four-port beam splitter on the -2nd, 0th and +2nd diffraction orders of NPGM with the dependence of wavelength, polarization and momentum of light, which is completely independent of the previous BS based on subspace I. 

Taking the BS process shown in Fig. 2C as an example, we can get the specific description $M_2$ for the subspace II through commercial COMSOL Multiphysics after excluding the propagation phase. It can be described as 
\begin{equation}
	\setlength{\arraycolsep}{2.5pt}
	\renewcommand{\arraystretch}{1.5}
	{\left( {\begin{array}{*{20}{c}}
				{|R\rangle_{ + 2}^d}\\
				{|L\rangle_0^d}\\
				{|R\rangle_0^a}\\
				{|L\rangle_{ + 2}^a}
		\end{array}} \right)_{out}} = \left( {\begin{array}{*{20}{c}}
			{0.44{e^{ - i \times 27^\circ }}}&{0.57{e^{ - i \times 64^\circ }}}&{0.48{e^{i \times 91^\circ }}}&{0.50{e^{ - i \times 44^\circ }}}\\
			{0.57{e^{ - i \times 63^\circ }}}&{0.45{e^{ - i \times 60^\circ }}}&{0.49{e^{ - i \times 96^\circ }}}&{0.48{e^{i \times 113^\circ }}}\\
			{0.48{e^{i \times 92^\circ }}}&{0.49{e^{ - i \times 96^\circ }}}&{0.59{e^{ - i \times 115^\circ }}}&{0.42{e^{ - i \times 95^\circ }}}\\
			{0.50{e^{ - i \times 44^\circ }}}&{0.48{e^{i \times 111^\circ }}}&{0.41{e^{ - i \times 96^\circ }}}&{0.59{e^{ - i \times 41^\circ }}}
	\end{array}} \right){\left( {\begin{array}{*{20}{c}}
				{|L\rangle_{ - 2}^d}\\
				{|R\rangle_0^d}\\
				{|L\rangle_0^a}\\
				{|R\rangle_{ - 2}^a}
		\end{array}} \right)_{in}}.
\end{equation}
Also, for the opposite BS process of Fig. 2C, whose input modes are $|R \rangle_{ + 2}^d,|L\rangle_0^d,|R\rangle_0^a,|L\rangle_{ + 2}^a$ and the output modes are $|L\rangle_{ - 2}^d,|R\rangle_0^d,|L\rangle_0^a,|R\rangle_{ - 2}^a$, the transformation matrix is the same as $M_2$. 

For subspace III shown in Fig. 2A, it corresponds to the most common conditions of eight-dimensional subspaces. Taking the subspace located at -3rd, -1st, +1st and +3st diffraction orders of NPGM as an example [Fig. 2D], it involves eight MPMs at eight different paths with the working wavelength of 1519.1 nm, including $|L\rangle_{ - 3}^d $, $|R \rangle_{ - 1}^d$, $|L\rangle_{ - 1}^a$, $|R\rangle_{ - 3}^a$, $|R\rangle_{ + 3}^d$, $|L\rangle_{ + 1}^d$, $|R\rangle_{ + 1}^a$, $|L\rangle_{ + 3}^a$. Similar to the condition of subspace II, these eight modes can be divided into four input modes and four output modes. If we choose $|L\rangle_{ - 3}^d $, $|R\rangle_{ - 1}^d$, $|L\rangle_{ - 1}^a$, $|R\rangle_{ - 3}^a$ as input modes, then the corresponding output modes are the combination of the other four modes [Fig. 2D]. Hereby, through these eight MPMs, we can get another four-port BS at another wavelength. The transformation matrix is similar to $M_2$ in form. Similarly, we can get a series of four-port BS from other subspcaes on NPGM, which has different working wavelengths and MPMs. These classical multiport BS can be extended to the quantum region for the  manipulation of quantum states.


\subsection*{Near-symmetric four-port BS for quantum light}
In this section, we will take the BS shown in Fig. 2C as an example to discuss the quantum BS on the NPGM. Firstly, all the eight MPMs can be denoted by annihilation operators. For the BS process shown in Fig. 2C, we label the input modes $|L\rangle_{-2}^d,|R\rangle_0^d,|L\rangle_0^a,|R\rangle_{ - 2}^a$ as ${\hat a_0},{\hat a_1},{\hat a_2},{\hat a_3}$ and the output modes $|R\rangle_{ + 2}^d,|L\rangle_0^d,|R\rangle_0^a,|L\rangle_{ + 2}^a$ as ${\hat b_0},{\hat b_1},{\hat b_2},{\hat b_3}$, respectively. These operators should satisfy Bosonic commutation relation $[{\hat a_i},{\hat a_j}]=[{\hat b_i},{\hat b_j}]=[\hat a_i^\dag ,\hat a_j^\dag ] = [\hat b_i^\dag ,\hat b_j^\dag ]=[{\hat a_i},\hat b_j^\dag]=0$ and ${\rm{[}}{\hat a_{i,}}\hat a_j^\dag {\rm{] = [}}{\hat b_{i,}}\hat b_j^\dag {\rm{] = }}{\delta _{ij}}$. Then the quantum BS can be represented by the transformation of operators. That is, after passing through NPGM, the operators $\hat a_i$ are transformed into the combination of operators $\hat b_i$ without the introduction of any vacuum operators [Fig. 3].
\begin{figure}[H]
	\centering
	\includegraphics[width=0.5\textwidth]{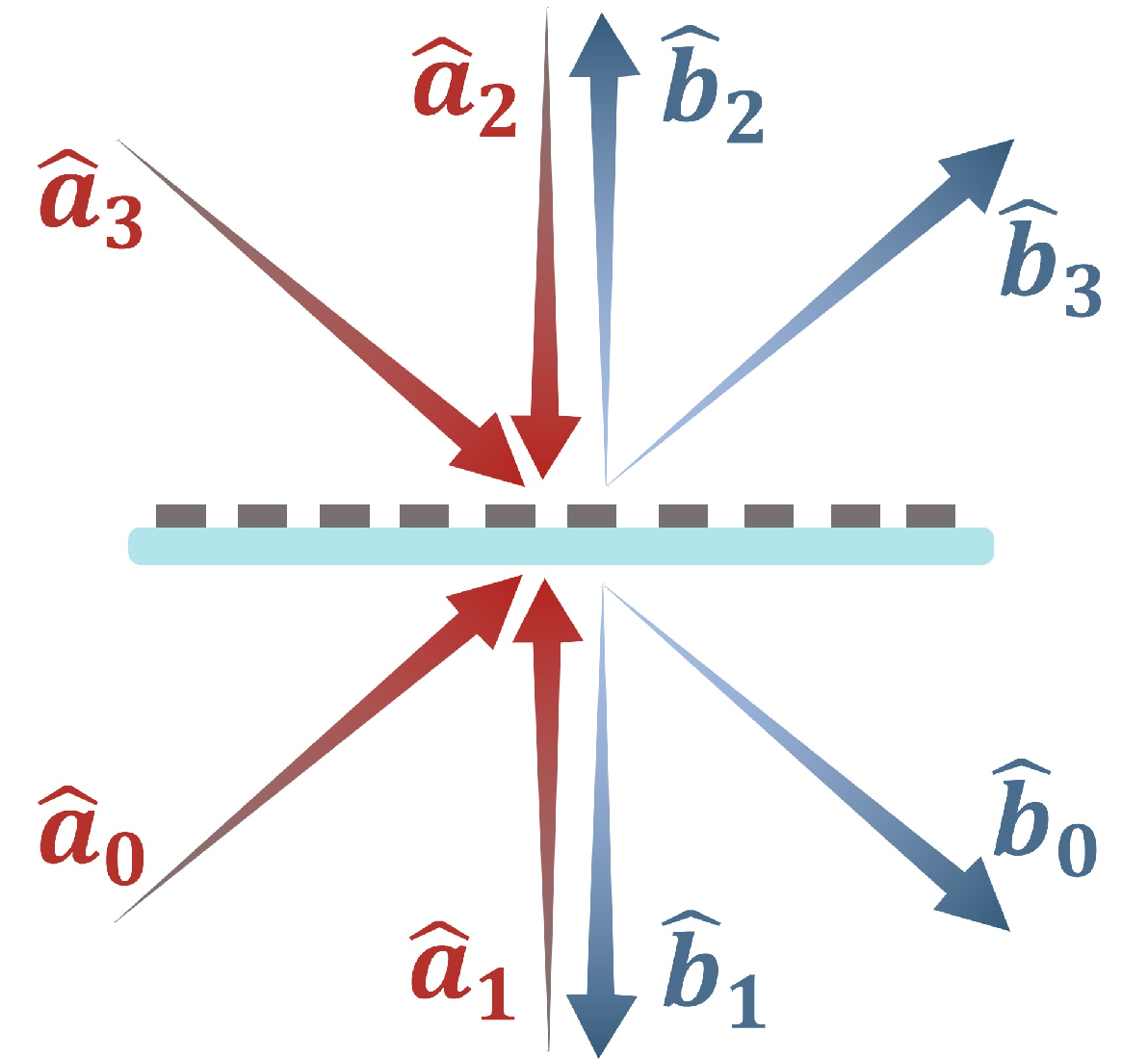}
	\caption{Four-port BS for quantum light.}	
\end{figure}

For the quantum BS shown in Fig. 3, we can get its specific transformation matrix through quantizing the above classical transformation matrix $M_2$. Due to the unitary of scattering matrix, we can write out the quantum scattering matrix $QM_2$ directly, which can be expressed as
\begin{equation}
	\renewcommand{\arraystretch}{1.5}
	\left( {\begin{array}{*{20}{c}}
			{{{\hat b}_0}}\\
			{{{\hat b}_1}}\\
			{{{\hat b}_2}}\\
			{{{\hat b}_3}}
	\end{array}} \right) = \left( {\begin{array}{*{20}{c}}
				{0.44{e^{ - i \times 27^\circ }}}&{0.57{e^{ - i \times 64^\circ }}}&{0.48{e^{i \times 91^\circ }}}&{0.50{e^{ - i \times 44^\circ }}}\\
			{0.57{e^{ - i \times 63^\circ }}}&{0.45{e^{ - i \times 60^\circ }}}&{0.49{e^{ - i \times 96^\circ }}}&{0.48{e^{i \times 113^\circ }}}\\
			{0.48{e^{i \times 92^\circ }}}&{0.49{e^{ - i \times 96^\circ }}}&{0.59{e^{ - i \times 115^\circ }}}&{0.42{e^{ - i \times 95^\circ }}}\\
			{0.50{e^{ - i \times 44^\circ }}}&{0.48{e^{i \times 111^\circ }}}&{0.41{e^{ - i \times 96^\circ }}}&{0.59{e^{ - i \times 41^\circ }}}
	\end{array}} \right)\left( {\begin{array}{*{20}{c}}
			{{{\hat a}_0}}\\
			{{{\hat a}_1}}\\
			{{{\hat a}_2}}\\
			{{{\hat a}_3}}
	\end{array}} \right).
\end{equation}
Based on the above quantum transformation matrix, we can obtain the effective Hamiltonian and the corresponding time evolution operator
\begin{equation}
	\begin{array}{l}
		{\hat H_{eff}} = {\hbar }\sum\limits_{m = 0}^{m = 3} {\sum\limits_{n = 0}^{n = 3} {{A_{m,n}}\hat a_m^\dag } } {\hat a_n},\\[5mm]
		\hat S(t) = \exp [ - \frac{i}{\hbar }{\hat H_{eff}}t],
	\end{array}
	\end{equation}
in which $A_{m,n}$ is the coupling coefficient among creation and annihilation operators (See Supplementary Materials S2). Under Heisenberg picture, the transformation relation among input and output operators is ${\hat b_i} = {\hat S^\dag }(t){\rm{ }}{\hat a_i}{\rm{ }}\hat S(t)$. Therefore, the transformation of any quantum states in this MPM subspace can be obtained. By utilizing this near-symmetric quantum multiport BS, we can realize the flexible and effective generation and manipulation of quantum states, especially the NOON states.
\subsection*{High-photon NOON states preparation based on near-symmetric four-port BS}
\begin{figure}[H]
	\centering
	\includegraphics[width=\textwidth]{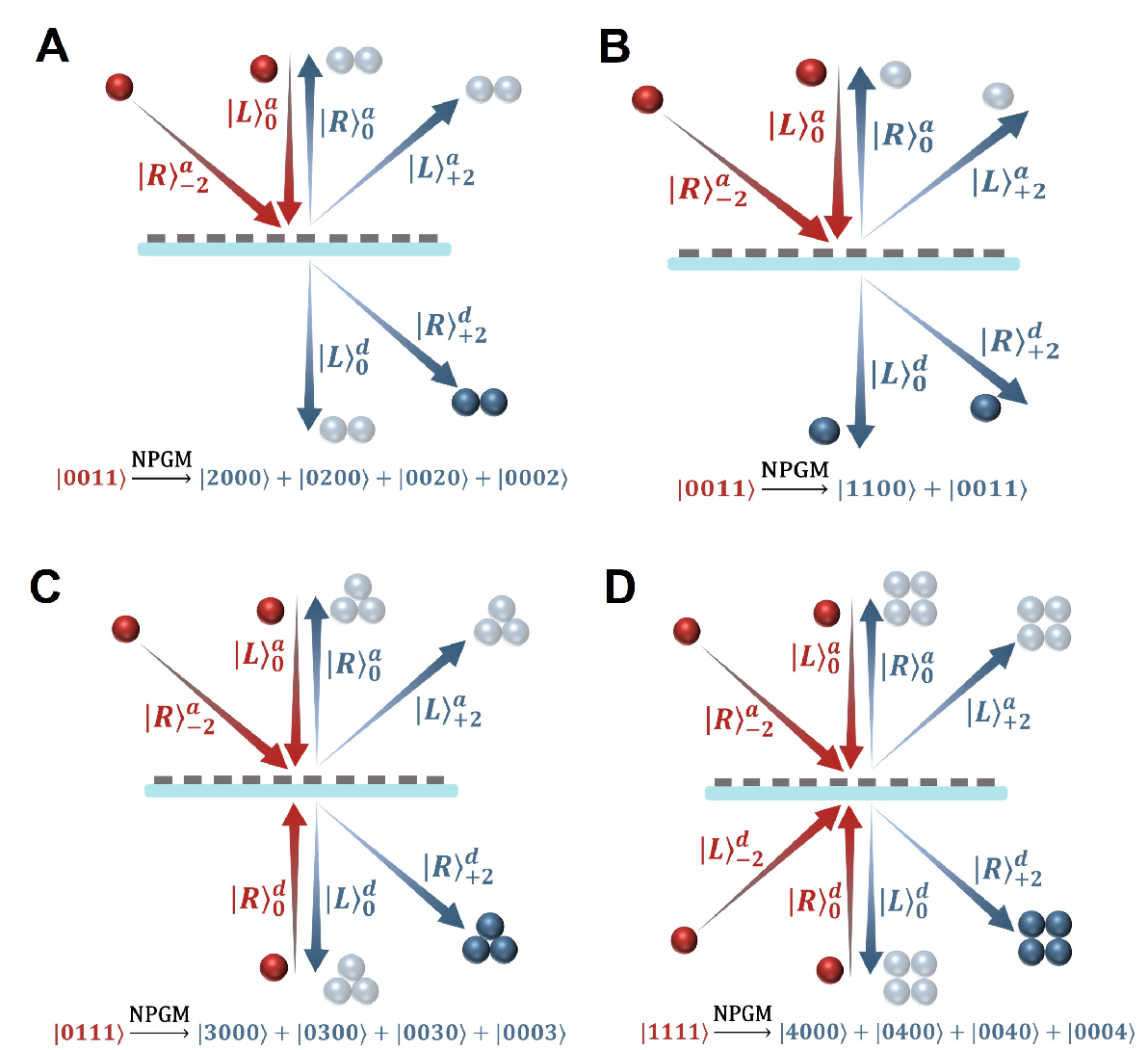}
	\caption{The preparation of high-photon NOON states through near-symmetric four-port quantum BS. (A) The schematic of the generation of four-mode two-photon  NOON state. (B) The schematic of path-entangled state during the process (A). (C) The schematic of the generation of four-mode three-photon  NOON state. (D) The schematic of the generation of four-mode four-photon NOON state.}	
\end{figure}

Through multiphoton interference, we achieve the preparation of four-mode high-photon NOON states with high success probability and fidelity. Firstly, we will discuss the simplest two-photon NOON state. Typicallly, two-mode two-photon NOON state can be deterministically generated through the interference of two identical photons by a 50/50 beam splitter. Instead, through two-photon interference on the NPGM, we can prepare four-mode two-photon NOON states with 50$\%$ success rate, as shown in Fig. 4A. The input state is 
${\left| \psi  \right\rangle _{in}} = \left| {0,0,1,1} \right\rangle  = \hat a_2^\dag \hat a_3^\dag \left| {0,0,0,0} \right\rangle $, then the output state ${\left| \psi  \right\rangle _{out}} = \hat S(t){\left| \psi  \right\rangle _{in}}$ can be expressed as  
\begin{align}
		{\left| \psi  \right\rangle _{out}} 
		&= 0.339{e^{i \times 48^\circ }}\left| {2,0,0,0} \right\rangle  + 0.333{e^{i \times 15^\circ }}\left| {0,2,0,0} \right\rangle  + 0.342{e^{i \times 149^\circ }}\left| {0,0,2,0} \right\rangle  \nonumber\\ 
		&+	0.350{e^{ - i \times 136^\circ }}\left| {0,0,0,2} \right\rangle   + 0.470{e^{ - i \times 148^\circ }}\left| {1,1,0,0} \right\rangle  \nonumber\\
		&+ 0.499{e^{ - i \times 167^\circ }}\left| {0,0,1,1} \right\rangle 
		+ 0.143{e^{ - i \times 124^\circ }}\left| {1,0,1,0} \right\rangle \nonumber\\
		&+ 0.085{e^{i \times 77^\circ }}\left| {1,0,0,1} \right\rangle  + 0.089{e^{i \times 14^\circ }}\left| {0,1,1,0} \right\rangle  + 0.143{e^{ - i \times 97^\circ }}\left| {0,1,0,1} \right\rangle .
\end{align}
From the output state, we can find the bunching effect occurs significantly, leading to the appearance of NOON state and disappearance of part of direct product states, for example, $| {1,0,0,1} \rangle $  state. Combining with coincidence measurement and phase shift, we can get four-mode two-photon NOON state with the success rate close to 50$\%$ and fidelity 99.9$\%$, which can be normalized as 
\begin{equation}
	\left| \psi  \right\rangle _{out}^{NOON} = 0.497\left| {2,0,0,0} \right\rangle  + 0.488\left| {0,2,0,0} \right\rangle  + 0.501\left| {0,0,2,0} \right\rangle  + 0.513\left| {0,0,0,2} \right\rangle .
\end{equation}
Besides, under this input condition, two-photon path-entangled state at different paths is generated simultaneously. From Eq.(7), by changing the methods for post-selection and phase shift and ignoring small probability state, we can acquire the normalized path-entangled state which can be expressed as 
\begin{equation}
	\left| \psi  \right\rangle _{out}^{Entangle} = 0.686\left| {1,1,0,0} \right\rangle  + 0.728\left| {0,0,1,1} \right\rangle 
\end{equation}
which means the two input photons are both reflected or transmitted by the NPGM and the two output photons are in the same side of NPGM, that is the air side or the glass side, as shown in Fig. 4B. Therefore,  the input two photons are converted into four-mode NOON state with 50$\%$ probability and entangled state with 50$\%$ probability through NPGM, realizing the multiplexing of   quantum state preparation on the NPGM.

Apart from two-photon NOON state, it has been proposed that multiphoton NOON states can be efficiently prepared by intensity-symmetric multiport beam splitters ~\cite{45}. Here, we demonstrate that our equivalent near-symmetric four-port beam splitter can generate three-photon and four-photon NOON states through multiphoton interference. For example, if the input state is ${\left| \psi  \right\rangle _{in}} = \left| {0,1,1,1} \right\rangle  = \hat a_1^\dag \hat a_2^\dag \hat a_3^\dag \left| {0,0,0,0} \right\rangle $, the NOON state component in the output state will be 
\begin{align}
		\left| \psi  \right\rangle _{out}^{NOON} &= 0.335{e^{-i \times 15^\circ }}\left| {3,0,0,0} \right\rangle  + 0.259{e^{ - i \times 45^\circ }}\left| {0,3,0,0} \right\rangle \nonumber\\
		&+ 0.290{e^{i \times 53^\circ }}\left| {0,0,3,0} \right\rangle  + 0.291{e^{ - i \times 23^\circ }}\left| {0,0,0,3} \right\rangle .
\end{align}
 Combined with phase shift for each NOON state component, we can get three-photon NOON state with the success rate 34.8$\%$ and fidelity 99.2$\%$, which can be normalized as
 \begin{equation}
 	\left| \psi  \right\rangle _{out}^{NOON} = 0.568\left| {3,0,0,0} \right\rangle  + 0.439\left| {0,3,0,0} \right\rangle  + 0.492\left| {0,0,3,0} \right\rangle  + 0.493\left| {0,0,0,3} \right\rangle ,
 \end{equation}
 as shown in Fig. 4C. It needs to be noted that the NOON states we prepare are encoded not only in the path modes, but also in the polarization modes. In other words, it is a kind of maximally three-photon path-polarization entangled state.
 
 Similarly, NPGM can also support four-photon NOON states generation from initially non-entangled state. If we input four-photon direct product state, just like ${\left| \psi  \right\rangle _{in}} = \left| {1,1,1,1} \right\rangle  = \hat a_0^\dag \hat a_1^\dag \hat a_2^\dag \hat a_3^\dag \left| {0,0,0,0} \right\rangle $, after post-selection, the NOON state component in the output state has the form like 
 \begin{align}
 		\left| \psi  \right\rangle _{out}^{NOON} &= 0.295{e^{ - i \times 42^\circ }}\left| {4,0,0,0} \right\rangle  + 0.296{e^{ - i \times 109^\circ }}\left| {0,4,0,0} \right\rangle \nonumber\\
 		&+ 0.279{e^{i \times 144^\circ }}\left| {0,0,4,0} \right\rangle  + 0.291{e^{ - i \times 67^\circ }}\left| {0,0,0,4} \right\rangle ).
 \end{align} 
 Combined with phase shift operation, we can acquire four-mode four-photon NOON state  with success rate 33.7$\%$ and fidelity close to 100$\%$, which can be normalized as
 \begin{equation}
 	\left| \psi  \right\rangle _{out}^{NOON} = 0.508\left| {4,0,0,0} \right\rangle  + 0.510\left| {0,4,0,0} \right\rangle  + 0.481\left| {0,0,4,0} \right\rangle  + 0.501\left| {0,0,0,4} \right\rangle , \end{equation}
 as shown in Fig. 4D. Therefore, we can prepare three-photon and four-photon NOON states via NGPM with the success rate higher than 1/3 and fidelity close to 100$\%$. Compared with previous scheme for preparing the NOON state, our scheme achieves multiphoton NOON state preparation with higher success rate and fidelity through a simpler and more integrated system, which can be applied in the quantum precision measurement.

\section*{Discussion}
This work demonstrates that NPGM can function as a series of near-symmetric four-port BS. The splitting ratio and port number of BS are jointly constrained by the single nonlocal mode and linear phase gradient, which restricts its scalable and flexible applications in the quantum state work and optical network. To overcome these two limitations, we can achieve asymmetric multiport BS with linear or elliptical polarization working basis through the coupling of multiple nonlocal modes~\cite{46} or the introduction of other phase gradient, just like exceptional topological phase~\cite{47}. Accordingly, the flexibility and functionality of the BS on the NPGM will be stronger.

On the basis of near-symmetric four-port BS on the NPGM, we discuss the preparation of up to four-photon NOON state. In fact, this scheme can be extended to the condition of more photons NOON states with relatively high success rate and fidelity by increasing incident photons,like we can get eight-photon NOON state by inputting two photons from four input ports, respectively. Moreover, the concentrated distribution of  photons at the incident ports will reduce the success rate and fidelity of NOON state preparation, which suggests that as many ports as possible in NOON state preparation through NPGM (see Supplementary Materials S3). 
	
In summary, we demonstrate that the NPGM can be regraded as a set of equivalent independent four-port beam splitters under resonance. These equivalent beam splitters are selective in the wavelength, polarization and momentum, which are constructed on the basis of a series of independent MPM subspaces. Taking this advantage, we prepare high-photon NOON states on the metasurface based on multiphoton quantum interference with high success probability and fidelity, for example, four-mode four-photon NOON state is obtained from four-photon direct product state with success rate 33.7$\%$ and fidelity close to 100$\%$. Combined with the momentum selectivity of nonlcoal modes and polarization-dependence of geometric phase, our results indicate the vast potential of NPGM in the multiphoton interference and quantum state work. Moreover, we provide a superior method for the efficient and integrated preparation of  high-photon NOON states which can pave the way for development of quantum precision measurement and quantum sensing.


\section*{Acknowledgments}
This work is supported by the National Natural Science Foundation of China under Grants No. 11974032 and the Innovation Program for Quantum Science and Technology under Grant No. 2021ZD0301500

\paragraph*{Conflict of interest.}The authors declare no competing interests.

\paragraph*{Data availability.}All data needed to evaluate the conclusions in the paper are present in the paper and/or the Supplementary Materials.

\section*{Supplementary materials}
\noindent
Supplementary Text S1 to S3\\
Figs. S1 to S2




\clearpage



\begin{thebibliography}{99}
	
	\bibitem{1} N. Yu, P. Genevet, M. A. Kats, F. Aieta, J.-P. Tetienne, F. Capasso, Z. Gaburro, Light Propagation with Phase Discontinuities: Generalized Laws of Reflection and Refraction. {\it Science} \textbf{334}, 333-337 (2011)
	
	\bibitem{2} H.-T. Chen, A. J. Taylor, N. Yu, A review of metasurfaces: physics and applications.  {\it Rep. Prog. Phys.} \textbf{79}, 076401 (2016)
	
	\bibitem{3} A. S. Solntsev, G. S. Agarwal, Y. S. Kivshar, Metasurfaces for quantum photonics.  {\it Nat. Photonics} \textbf{15}, 327-336 (2021)
	
	\bibitem{4} A. S. M. Khorasaninejad, K. B. Crozier, Silicon nanofin grating as a miniature chirality-distinguishing beam-splitter.  {\it Nat. Commun.} \textbf{5}, 5386 (2014)
	
	\bibitem{5} K. Wang, J. G. Titchener, S. S. Kruk, L. Xu, H.-P. Chung, M. Parry, I. I. Kravchenko, Y.-H. Chen, A. S. Solntsev, Y. S. Kivshar, D. N. Neshev, A. A. Sukhorukov, Quantum metasurface for multiphoton interference and state reconstruction.  {\it Science} \textbf{361}, 1104-1108 (2018)
	
	\bibitem{6} P. Georgi, M. Massaro, K.-H. Luo, B. Sain, N. Montaut, H. Herrmann, T. Weiss, G. Li, C. Silberhorn, T. Zentgraf, Metasurface interferometry toward quantum sensors.  {\it Light Sci. Appl.} \textbf{8}, 70 (2019)
	
	\bibitem{7} Y.-J. Gao, X. Xiong, Z. Wang, F. Chen, R.-W. Peng, M. Wang, Simultaneous Generation of Arbitrary Assembly of Polarization States with Geometrical-Scaling-Induced Phase Modulation.  {\it Phys. Rev. X } \textbf{10}, 031035 (2020)
	
	\bibitem{8} Y.-J. Gao, Z. Wang, Y. Jiang, R.-W. Peng, Z.-Y. Wang, D.-X. Qi, R.-H. Fan, W.-J. Tang, M. Wang, Multichannel Distribution and Transformation of Entangled Photons with Dielectric Metasurfaces.   {\it Phys. Rev. Lett. } \textbf{129}, 023601 (2022)
	
	\bibitem{9} Y. Li, X. Fan, X. Guo, Y. Zhang, S. Liu, B. Wei, D. Wen, P. Li, J. Zhao, Metasurface for oscillatory spin splitting along the optical path.  {\it Photonics Res. } \textbf{10}, B7-B13 (2022)
	
	\bibitem{10} 	Q. Liu, Z. Liu, X. Ma, J. Deng, C. Zhang, Z. Chen, A. Nemati, S. K. Ng, S. Gorelik, S. L. Teo, R. Ji, M. Zhao, L. V. Gonzaga, H. Liu, F. Yue, S. Yu, Y. Luo, Q. Wang, Incomplete Phase Metasurface for Wavefront Reconstruction.  {\it ACS Photonics  } \textbf{10}, 2563-2569 (2023)
	
	\bibitem{11} 	Y. Ke, Y. Bian, Q. Tang, J. Tian, L. Zeng, Y. Chen, X. Zhou, Rotational photonic spin Hall effect.  {\it Nanophotonics  } \textbf{12}, 4361-4373 (2023)
	
	\bibitem{12} 	Z. Li, E. Palacios, S. Butun, K. Aydin, Visible-Frequency Metasurfaces for Broadband Anomalous Reflection and High-Efficiency Spectrum Splitting.  {\it Nano Lett.  } \textbf{15}, 1615-1621 (2015)
	
	\bibitem{13} 	Z. Li, E. Palacios, S. Butun, K. Aydin, Ultrawide Angle, Directional Spectrum Splitting with Visible-Frequency Versatile Metasurfaces.   {\it Adv. Opt. Mater.  } \textbf{4}, 953-958 (2016)
	
	\bibitem{14} 	Z. Su, X. Chen, J. Yin, X. Zhao, Graphene-based terahertz metasurface with tunable spectrum splitting.    {\it Opt. Lett.  } \textbf{41}, 3799-3802 (2016)
	
	\bibitem{15} 	A. Silva, F. Monticone, G. Castaldi, V. Galdi, A. Alù, N. Engheta, Performing Mathematical Operations with Metamaterials.   {\it Science  } \textbf{343}, 160-163 (2014)
	
	\bibitem{16} Y. Yang, W. Wang, A. Boulesbaa, I. I. Kravchenko, D. P. Briggs, A. Puretzky, D. Geohegan, J. Valentine, Nonlinear Fano-Resonant Dielectric Metasurfaces.   {\it Nano Lett.  } \textbf{15}, 7388-7393 (2015)
	
	\bibitem{17} H. Kwon, D. Sounas, A. Cordaro, A. Polman, A. Alù, Nonlocal Metasurfaces for Optical Signal Processing.   {\it Phys. Rev. Lett.  } \textbf{121}, 173004 (2018)
	
	\bibitem{18} J.-H. Song, J. van de Groep, S. J. Kim, M. L. Brongersma, Non-local metasurfaces for spectrally decoupled wavefront manipulation and eye tracking.  {\it Nat. Nanotechnol.  } \textbf{16}, 1224-1230 (2021)
	
	\bibitem{19} A. Overvig, A. Alù, Diffractive Nonlocal Metasurfaces. {\it Laser Photonics Rev.  } \textbf{16}, 2100633 (2022)
	
	\bibitem{20} K. Shastri, F. Monticone, Nonlocal flat optics. {\it Nat. Photonics  } \textbf{17}, 36-47 (2023)
	
	\bibitem{21} R. Kolkowski, T. K. Hakala, A. Shevchenko, M. J. Huttunen, Nonlinear nonlocal metasurfaces. {\it Appl. Phys. Lett.  } \textbf{122}, 160502 (2023)
	
	\bibitem{22} C. W. Hsu, B. Zhen, A. D. Stone, J. D. Joannopoulos, M. Soljačić, Bound states in the continuum. {\it Nat. Rev. Mater.   } \textbf{1}, 1-13 (2016)
	
	\bibitem{23} K. Koshelev, S. Lepeshov, M. Liu, A. Bogdanov, Y. Kivshar, Asymmetric Metasurfaces with High-Q Resonances Governed by Bound States in the Continuum.  {\it Phys. Rev. Lett.} \textbf{121}, 193903 (2018)
	
	\bibitem{24} T. Santiago-Cruz, S. D. Gennaro, O. Mitrofanov, S. Addamane, J. Reno, I. Brener, M. V. Chekhova, Resonant metasurfaces for generating complex quantum states. {\it Science} \textbf{377}, 991-995 (2022)
	
	\bibitem{25} 	J. Zhang, J. Ma, M. Parry, M. Cai, R. Camacho-Morales, L. Xu, D. N. Neshev, A. A. Sukhorukov, Spatially entangled photon pairs from lithium niobate nonlocal metasurfaces. {\it Sci. Adv.} \textbf{8}, eabq4240 (2022)
	
	\bibitem{26} 	A. Aigner, A. Tittl, J. Wang, T. Weber, Y. Kivshar, S. A. Maier, H. Ren, Plasmonic bound states in the continuum to tailor light-matter coupling. {\it Sci. Adv.} \textbf{8}, eadd4816 (2022)
	
	\bibitem{27} 	M. Kang, T. Liu, C. T. Chan, M. Xiao, Applications of bound states in the continuum in photonics.  {\it Nat. Rev. Phys.} \textbf{5}, 659-678 (2023)
	
	\bibitem{28} A. Tittl, A. Leitis, M. Liu, F. Yesilkoy, D.-Y. Choi, D. N. Neshev, Y. S. Kivshar, H. Altug, Imaging-based molecular barcoding with pixelated dielectric metasurfaces.   {\it Science } \textbf{360}, 1105-1109 (2018)
	
	\bibitem{29} 	F. Yesilkoy, E. R. Arvelo, Y. Jahani, M. Liu, A. Tittl, V. Cevher, Y. Kivshar, H. Altug, Ultrasensitive hyperspectral imaging and biodetection enabled by dielectric metasurfaces.    {\it Nat. Photonics } \textbf{13}, 390-396 (2019)
	
	\bibitem{30} 	A. Leitis, A. Tittl, M. Liu, B. H. Lee, M. B. Gu, Y. S. Kivshar, H. Altug, Angle-multiplexed all-dielectric metasurfaces for broadband molecular fingerprint retrieval.    {\it Sci. Adv. } \textbf{5}, eaaw2871 (2019)
	
	\bibitem{31} 	A. C. Overvig, S. C. Malek, M. J. Carter, S. Shrestha, N. Yu, Selection rules for quasibound states in the continuum.   {\it Phys. Rev. B. } \textbf{102}, 035434 (2020)
	
	\bibitem{32} 	A. C. Overvig, S. C. Malek, N. Yu, Multifunctional Nonlocal Metasurfaces.  {\it Phys. Rev. Lett.  } \textbf{125}, 017402 (2020)
	
	\bibitem{33} 	A. Overvig, A. Alù, Wavefront-selective Fano resonant metasurfaces.   {\it  Adv. Photonics } \textbf{3}, 026002 (2021)
	
	\bibitem{34} 	A. Overvig, N. Yu, A. Alù, Chiral Quasi-Bound States in the Continuum.   {\it  Phys. Rev. Lett. } \textbf{126}, 073001 (2021)
	
	\bibitem{35} 	S. C. Malek, A. C. Overvig, A. Alù, N. Yu, Multifunctional resonant wavefront-shaping meta-optics based on multilayer and multi-perturbation nonlocal metasurfaces.  {\it  Light Sci. Appl.  } \textbf{11}, 246 (2022)
	
	\bibitem{36} 	G. Xu, A. Overvig, Y. Kasahara, E. Martini, S. Maci, A. Alù, Arbitrary aperture synthesis with nonlocal leaky-wave metasurface antennas.   {\it Nat. Commun.   } \textbf{14}, 4380 (2023)
	
	\bibitem{37} T. Liu, D. Zhang, W. Liu, T. Yu, F. Wu, S. Xiao, L. Huang, A. E. Miroshnichenko, Phase-change nonlocal metasurfaces for dynamic wave-front manipulation.  {\it Phys. Rev. Appl. } \textbf{21}, 044004 (2024)
	
	\bibitem{38} 	A. Overvig, S. A. Mann, A. Alù, Spatio-temporal coupled mode theory for nonlocal metasurfaces. {\it Light Sci. Appl.  } \textbf{13}, 28 (2024)
	
	\bibitem{39} 	J. J. . Bollinger, W. M. Itano, D. J. Wineland, D. J. Heinzen, Optimal frequency measurements with maximally correlated states.  {\it Phys. Rev. A } \textbf{54}, R4649–R4652 (1996)
	
	\bibitem{40}    V. Giovannetti, S. Lloyd, L. Maccone, Advances in quantum metrology.  {\it Nat. Photonics  } \textbf{5}, 222–229 (2011)
	
	\bibitem{41}    J. Wang, G. S. Agarwal, Quantum Fisher information bounds on precision limits of circular dichroism.   {\it Phys. Rev. A  } \textbf{104}, 062613 (2021)
	
	\bibitem{42}    	I. Afek, O. Ambar, Y. Silberberg, High-NOON States by Mixing Quantum and Classical Light. {\it Science  } \textbf{328}, 879-881 (2010)
	
	\bibitem{43}    		M. W. Mitchell, J. S. Lundeen, A. M. Steinberg, Super-resolving phase measurements with a multiphoton entangled state.  {\it Nature  } \textbf{429}, 161-164 (2004)
	
	\bibitem{44}    J. C. F. Matthews, A. Politi, D. Bonneau, J. L. O’Brien, Heralding Two-Photon and Four-Photon Path Entanglement on a Chip. {\it Phys. Rev. Lett.  } \textbf{107}, 163602 (2011)
	
	\bibitem{45}   G. J. Pryde, A. G. White, Creation of maximally entangled photon-number states using optical fiber multiports. {\it Phys. Rev. A } \textbf{68}, 052315 (2002)
	
	\bibitem{46}    Z. Su, Y. Yang, B. Xiong, R. Zhao, Y. Wang, L. Huang, Planar Chiral Metasurface Based on Coupling Quasi-Bound States in the Continuum. {\it Adv. Opt. Mater.  } \textbf{12}, 2303195 (2024)
	
	\bibitem{47}    Q. Song, M. Odeh, J. Zúñiga-Pérez, B. Kanté, P. Genevet, Plasmonic topological metasurface by encircling an exceptional point. {\it Science } \textbf{373}, 1133–1137 (2021)
\end{thebibliography}
\end{document}